\documentclass{amsart}
\usepackage{amsmath,amssymb,dsfont,mathtools,amsthm,enumerate,lineno}
  \usepackage{paralist}
  \usepackage{graphics} 
  \usepackage{epsfig} 
\usepackage{graphicx}

\mathtoolsset{showonlyrefs}

 \usepackage[colorlinks=true]{hyperref}

   \hypersetup{urlcolor=blue, citecolor=red}

\usepackage{hyperref}

\usepackage{siunitx}

\numberwithin{equation}{section}
\allowdisplaybreaks[4]
\newcommand{\xqedhere}[2]{%
	\rlap{\hbox to#1{\hfil\llap{\ensuremath{#2}}}}}

\newtheorem{theorem}{Theorem}[section]

\newtheorem{lemma}[theorem]{Lemma}

\theoremstyle{definition}

\newcommand{\f}{2\Omega}
\newcommand{\z}{\bar{z}}
\newcommand{\abs}[1]{\left\vert#1\right\vert}
\renewcommand{\vec}[1]{\mathbf{#1}}
\newcommand{\re}[1]{\mathfrak{Re}\left\{#1\right\}}
\newcommand{\im}[1]{\mathfrak{Im}\left\{#1\right\}}
\newcommand{\dF}{F^{\prime}(z)}
\newcommand{\dG}{G^{\prime}(\bar{z})}
\renewcommand{\l}[1]{\lambda_{#1}}
\newcommand{\s}[1]{\sigma_{#1}}
\newcommand{\w}[1]{\omega_{#1}}

\title[Particle paths in equatorial flows]{Particle paths in equatorial flows}

\author{Tony Lyons}
\address{Department of Computing and Mathematics, Waterford Institute of Technology, Waterford, Ireland}
\subjclass{Primary: 76U60; Secondary: 76B15, 76B47.}
\keywords{Geophysical flows, 2-dimensional flows, particle trajectories, Lagrangian variables, vorticity.}

 \email{tlyons@wit.ie}

\begin{document}
\maketitle





\begin{abstract}
We investigate particle trajectories in equatorial flows with geophysical corrections caused by the earth's rotation. Particle trajectories in the flows are constructed using pairs of analytic functions defined over the labelling space used in the Lagrangian formalism. Several classes of flow are investigated, and the physical regime in which each is valid is determined using the pressure distribution function of the flow, while the vorticity distribution of each flow is also calculated and found to be effected by earth's rotation.
\end{abstract}

\section{Introduction}
We consider particle paths in two-dimensional geophysical flows in close proximity to the equator, where the so called $f$-plane approximation is reasonably accurate. The work is based on the Lagrangian description of hydrodynamics \cite{Ben2006}, which has been applied to a wide variety of geophysical flows in recent years. We do not restrict the flows to be irrotational, and as such we cannot directly use harmonic maps to describe the particle trajectories, a method which has been applied with great success to irrotational flows (see \cite{Con2011,MT2013} for instance).  Nevertheless, we may describe the flows in terms of pairs of analytic maps, defined over the labelling space used in the Lagrangian formalism, the important difference being that these are analytic functions of complex-conjugate variables, thus allowing us to incorporate vorticity in the flows.

Such a procedure for describing flows with vorticity was first introduced in \cite{CS2010} and in the current work we develop this further to allow for Coriolis forces near the equator, which arises due to the rotation of of the earth about its polar axis. In some cases the particle trajectories are not associated with free boundary flows, in that there are no streamlines which are subject to constant pressure. However, in one case we show that it is possible to recover a free-boundary flow in the form of trochoidal waves. Such trochoidal waves have been used extensively to model a wide variety of geophysical phenomena, (see \cite{Hen2015, Hen2016, Klu2018, Rod2018} for various applications of trochoidal waves in the $f$-plane approximation).

The analysis of hydrodynamic flows in terms of particle trajectories in the fluid has been a remarkably fruitful avenue of recent research, for instance, the description of particle paths in irrotational, inviscid fluids has been investigated in \cite{CV2008, CEV2008} where particle trajectories in linear waves were analysed, while fluid particle trajectories in fully nonlinear Stokes waves have been investigated in \cite{Con2006, Con2012a, CS2010a, Hen2006, Hen2008a, IM2014, Lyo2014, Mat2012, Qui2017} among others.

In the current frame work we do not initially assume the flows to be rotational, so many of the techniques used to analyse particle trajectories in irrotational flows are not applicable. However, the Lagrangian framework provides an alternative avenue for the analysis, a particularly appealing feature of this approach being that it may be applied to the fully nonlinear governing equations of hydrodynamics. In Section \ref{sec:governing} we present the governing equations describing geophysical flows near the equator as well as the boundary conditions required of the solutions of this system. In Section \ref{sec:lagrangian} we outline the construction of solutions to these governing equations in the Lagrangian frame work, whereby the particle position in the fluid domain is given by a diffeomorphism from the labelling space used to identify the individual particles of the flow. In Section \ref{sec:flows} we describe these diffeomorphisms using pairs of analytic functions defined over the complex plane, and determine three distinct analytic functions admissible as potential solutions. In the final part of the paper we construct explicit particle trajectories and examine under what conditions they may be physically realised. In this regard the pressure distribution will play a central role in determining where the flow may be valid. The effects of geophysical corrections are most apparent when we consider the vorticity within each flow as will be found in the following.

 \section{The governing equations}\label{sec:governing}
 The governing equations for 2-dimensional geophysical flows near the equator, incorporating the Coriolis force induced by earth's rotation, are conventionally written in the $f$-plane setting  as
 \begin{equation}\label{eq:euler}
 \begin{cases}
  \displaystyle u_{t} + uu_{x} + vu_{y} + \f v = -\frac{1}{\rho}P_{x} \\
 \displaystyle  v_{t} + uv_{x} + vv_{y}- \f u \!=\! -\frac{1}{\rho}P_{y} - g
 \end{cases}
 \end{equation}
where the $x$-axis is along the zonal direction and increases moving eastwards, while $y$ is the vertical coordinate above the surface of the earth, with $u$ and $v$ being the velocities along these axes. The hydrodynamic pressure is denoted by $P$, while the fluid density is denoted by $\rho$, which in the following we may reasonably set to the constant $\rho=\SI{1e3}{\kilo\gram\per\cubic\meter}$ and by re-scaling $\frac{1}{\rho}P\to P$ we can set $\rho=1$ without loss of generality . The Coriolis force components $-\f u$ and $\f v$ are a consequence of earth's rotation with the angular velocity of earth given by $\Omega=\SI{7.29e-5}{\radian\per\second}$, while $g=\SI{9.8}{\meter\per\square\second}$ is the acceleration due to gravity. Coupled with this equation is the equation of mass conservation in the flow $\rho_{t}+\nabla\cdot\left(\rho\vec{u}\right)=0$, which in the present regime becomes
\begin{equation}\label{eq:massconservation}
 u_{x}+v_{y}=0.
\end{equation}
In addition, the boundary conditions on the free surface $y=\eta(x,t)$ are given by
\begin{equation}\label{eq:surfacecondition}
 \begin{rcases}
 P=P_{atm} \\
 v=\eta_{t}+u\eta_x
 \end{rcases}\text{ on }z=\eta(x,t)
 \end{equation}
where $P_{atm}$ is the \emph{constant} atmospheric pressure exerted on the free surface. Meanwhile the condition that fluid motion cease at great depth is imposed by
\begin{equation}\label{eq:deepcondition}
 (u,v)\to(0,0)\text{ as }y=\to-\infty.
\end{equation}
In contrast to the typical change of direction common at mid-latitudes, where bending is common and gyres dominate the ocean flow (see \cite{CJ2017} for further discussion), the two-dimensionality of flows in the equatorial region occurs due to the change of sign of the Coriolis parameter across the equator producing an effective wave-guide which facilitates azimuthal flow propagation (cf. see \cite{CI2019}).

\section{The flow in Lagrangian variables}\label{sec:lagrangian}
In the Lagrangian formalism, the governing equation \eqref{eq:euler} simply becomes
\begin{subequations}\label{eq:eulerlagrange0}
 \begin{align}
  \ddot{x} + \f \dot{y} &= - P_{x} \\
  \ddot{y} - \f \dot{x} &= - P_{y} - g
 \end{align}
 \end{subequations}
where $\dot{x}=\frac{d x}{dt}=u(x(t),y(t))$ and $\dot{y}=\frac{dy}{dt}=v(x(t),y(t))$ are the velocity components of the particle located at $\left(x(t),y(t)\right)$ at the instant $t$.
We propose a class of solutions of the system \eqref{eq:euler}--\eqref{eq:deepcondition} of the form
\begin{equation}\label{eq:diffeomorphism}
\begin{aligned}
x(t) = f(a+ct,b) \\
y(t) = h(a+ct,b) \\
\end{aligned}
\end{equation}
where $(x(t),y(t))$ are the coordinates of an individual fluid particle at time $t$, with $(a,b)$ serving as the Lagrangian variables for the fluid particle in question. It will be seen that the parameter $b$ labels the fluid particle trajectory, while $a+ct$ denotes the position of the particle along its trajectory at the instant $t$, with $c$ being the wave-speed of the surface wave with respect to the $(x,y)$-coordinate system. The system \eqref{eq:eulerlagrange0} when written in terms of the Lagrangian variables $a$, $b$ becomes
 \begin{subequations}
 \begin{align}
 \label{eq:eulerlagrange1} c^2f_{aa} + \f ch_{a} &= - P_{x} \\
 \label{eq:eulerlagrange2} c^2h_{aa} - \f cf_{a} &= - P_{y} - g.
 \end{align}
 \end{subequations}
The Jacobian matrix of the coordinate map \eqref{eq:diffeomorphism} is given by
 \begin{equation}\label{eq:jacobianmatrix}
 \frac{\partial\left(x,y\right)}{\partial\left(a,b\right)} = \begin{bmatrix}
                                                               f_{a} & h_{a} \\
                                                               f_{b} & h_{b}
                                                             \end{bmatrix},
 \end{equation}
 with the corresponding Jacobian given by
 \begin{equation}\label{eq:jacobian}
   J=f_{a}h_{b}-f_{b}h_{a}.
 \end{equation}
 Inverting the Jacobian matrix in \eqref{eq:jacobianmatrix} to find $\left(\partial_{x},\partial_{y}\right)$ in terms of $\left(\partial_{a},\partial_{b}\right)$, one may confirm that $J_{t}=u_{x}+v_{y},$ in which case the mass conservation equation \eqref{eq:massconservation} ensures $J_{t}=0$, meaning volume is conserved along such flows. The Jacobian \eqref{eq:jacobian} is required to satisfy $J>0$ (although certain flows may allow $J=0$ for a single value of $b\in\mathbb{R}$ for instance trochoidal flows). Moreover, since $J=J(a+ct,b)$ and $J_{t}=0$, it follows at once that $J=\mu(b)$ for some function $\mu$.

 Left-multiplying the pressure gradient $\left(P_{x} \ P_{y}\right)^{\mathrm{T}}$ from \eqref{eq:eulerlagrange1}--\eqref{eq:eulerlagrange2} by the Jacobian matrix in \eqref{eq:jacobianmatrix} yields the pressure gradient with respect to the Lagrangian variables:
 \begin{subequations}
 \begin{align}
  \label{eq:pressuregradienta} P_{a} &= -c^2\left(f_{a}f_{aa} + h_{a}h_{aa}\right) - gh_a \\
  \label{eq:pressuregradientb} P_{b} &= -c^2\left(f_{b}f_{aa} + h_{b}h_{aa}\right) - gh_b + \f c \mu.
 \end{align}
 \end{subequations}
 Integrating \eqref{eq:pressuregradienta} we find
 \[P(a,b)= -\frac{c^2}{2}\left(f_{a}^2 + h_{a}^2\right) + gh + \nu(b),\]
 while equation \eqref{eq:pressuregradientb} gives
 \begin{equation}\label{eq:nuprime0}
 \nu^{\prime}(b) = c^{2} \left( f_{a}f_{ab} + h_{a}h_{ab} - f_{b}f_{aa} - h_{b}h_{aa} \right) - \f c\mu.
 \end{equation}

 \section{Flows from analytic maps}\label{sec:flows}
 The functions  $f$ and $g$ are chosen so that the complex map given by
 \begin{equation}\label{eq:complexmap}
 f(a,b) + ih(a,b) = F(z) + G(\z),\quad z = a + ib,
 \end{equation}
 is restricted by the condition  $F$ and $G$ are analytic functions and of $z$ and $\z = a - ib$ respectively. Since $F$ and $G$ chosen so that both are analytic, they may be written in the form
 \begin{equation}
 F(z) = \gamma(a,b) + i\delta(a,b), \quad G(\z) = \alpha(a,b) + i\beta(a,b)
 \end{equation}
 where $\left\{\alpha,\beta\right\}$ and $\left\{\gamma,\delta\right\}$ are harmonic conjugate pairs, both satisfying the Cauchy-Riemann equations
 \begin{equation}\label{eq:cauchyriemann}
 \begin{cases}
 \alpha_{a} = \beta_{b} \\
 \alpha_{b} = -\beta_{a}
 \end{cases} \qquad
 \begin{cases}
  \gamma_{a} = - \delta_{b} \\
  \gamma_{b} = \delta_{a}.
 \end{cases}
 \end{equation}
We now deduce from equations \eqref{eq:jacobian}, \eqref{eq:pressuregradientb} and \eqref{eq:cauchyriemann} that
 \begin{subequations}
 \begin{align}
 \label{eq:mu}\mu(b) &= \gamma_{a}^{2} + \gamma_{b}^2 - \alpha_{a}^2 - \alpha_{b}^2 \\
 \label{eq:nuprime}\nu^{\prime}(b) &= c^2\partial_{b}\left(\alpha_{a}^{2}+\alpha_{b}^{2}+\gamma_{a}^2+\gamma_{b}^2\right) + \f c\mu(b).
 \end{align}
 \end{subequations}
 The conditions \eqref{eq:mu}--\eqref{eq:nuprime} are assured if there exists a function $\xi(a)$ such that
 \begin{subequations}
 \begin{align}
 \label{eq:Fmunuxi}\abs{F^{\prime}(z)}^2  &= \gamma_{a}^{2} + \gamma_{b}^{2} = \frac{\mu(b)}{2} + \frac{1}{2c^2}\nu(b) + \frac{\Omega}{c} \int\mu(b) db + \xi(a) \\
 \label{eq:Gmunuxi}\abs{G^{\prime}(\z)}^2 &= \alpha_{a}^{2} + \alpha_{b}^{2} = -\frac{\mu(b)}{2} + \frac{1}{2c^2}\nu(b) + \frac{\Omega}{c} \int\mu(b)  db + \xi(a),
 \end{align}
 \end{subequations}
 from which we immediately deduce
 \begin{equation}
 \partial_{a}\partial_{b}\abs{F^{\prime}(z)}^{2} = \partial_{a}\partial_{b}\abs{G^{\prime}(\z)}^{2} = 0.
 \end{equation}

 \begin{lemma}\label{lem:Lemma1}\cite{CS2010}
 All analytic functions $F$ satisfying
 \begin{equation}\label{lem:condition}
 \partial_{a}\partial_{b}\abs{F^{\prime}(z)}^2=0,\qquad z=a+ib
 \end{equation}
 have the form
 \begin{equation}\label{lem:formsofF}
 F(z) = \begin{cases}
          \w{0} + \w{1}z + \w{2}z^2 \\
          \w{0} + \w{1}e^{kz} + \w{2}e^{-kz} \\
          \w{0} + \w{1}e^{ikz} + \w{2}e^{-ikz},
        \end{cases}
 \end{equation}
 where $k\in\mathbb{R}\backslash\{0\}$ and $\w{0},\w{1},\w{2}\in\mathbb{C}$ are arbitrary constants.
 \end{lemma}

We recall some important aspects of the proof of this lemma as presented in \cite{CS2010} and further related results in \cite{AC2012, CM2017, MT2019}. Let $F(z)$ be an analytic function such that $\partial_{a}\partial_{b}\abs{F^{\prime}(z)}^2=0$ and a region $\mathcal{R}\subseteq\mathbb{C}$ where ${F^{\prime}(z)}\neq0$ for all $z\in\mathcal{R}$. Since ${F^{\prime}(z)}\neq0$ in this region, it may be written as
 \begin{equation}\label{lem:fprime}
 F^{\prime}(z)=e^{i\left(\tilde{f}(a,b)+i\tilde{h}(a,b)\right)},
 \end{equation}
 where $\tilde{f}$ and $\tilde{h}$ are both harmonic. The condition \eqref{lem:condition} now yields
 \begin{equation}\label{lem:hab}
 \tilde{h}_{ab}=2\tilde{h}_{a}\tilde{h}_{b},
 \end{equation}
 whose solution is of the form
 \begin{equation}\label{lem:h=}
 \tilde{h}(a,b)=-\frac{1}{2}\ln\left[A(a)+B(b)\right],
 \end{equation}
 where $A(a)$ and $B(b)$ are arbitrary real valued functions such that $A(a)+B(b)>0$ for all $\left\{a, b: a+ib\in\mathcal{R}\right\}$.

 We require $\tilde{h}$ to be harmonic, in which case we must have
 \begin{equation}\label{lem:hharmonic}
 (A_{aa}+B_{bb})(A+B)-A_{a}^2-B_{b}^2=0,
 \end{equation}
 while applying $\partial_{a}$ and $\partial_{b}$ to \eqref{lem:hharmonic} we have
 \begin{equation}\label{lem:pdeAB}
 A_{a}B_{bbb} + B_{b}A_{aaa} = 0.
 \end{equation}
 Separation of variables then yields three classes of solutions given by:
 \begin{subequations}
 \begin{align}
 \label{lem:polynomialAB}
 &\begin{cases}
  A(a) = \l{0} + \l{1}a + \l{2}a^2\\
  B(b) = \s{0} + \s{1}b + \s{2}b^2
 \end{cases}\\
 \label{lem:trigAexpB}
 &\begin{cases}
  A(a) = \l{0} + \l{1}\cos(ka) + \l{2}\sin(ka)\\
  B(b) = \s{0} + \s{1}e^{kb} + \s{2}e^{-kb}
 \end{cases}\\
 \label{lem:expAtrigB}
 &\begin{cases}
  A(a) = \l{0} + \l{1}e^{ka} + \l{2}e^{-ka}\\
  B(b) = \s{0} + \s{1}\cos(kb) + \s{2}\sin(kb),
 \end{cases}
 \end{align}
 \end{subequations}
 where $\l{0}$, $\l{1}$, $\l{2}$, $\s{0}$, $\s{1}$, $\s{2}$ and $k\neq0$ are arbitrary real constants.

 We now consider the conditions each class of solution must satisfy given $\tilde{h}$ is real and harmonic:
 \begin{subequations}
 \begin{enumerate}[\bfseries i)]
 \item Applying condition \eqref{lem:hharmonic} to \eqref{lem:polynomialAB} we find that
 \begin{equation}\label{lem:polynomialconditions}
  \begin{aligned}
  & \l{1}=\l{2}=\s{1}=\s{2}=0 \\
  &  \text{ or } \\
  &  \l{2}=\s{2}>0 \text{ \& } 4\l{2}(\l{0}+\s{0})=\l{1}^2+\s{1}^2.
  \end{aligned}
 \end{equation}

 \item Substituting \eqref{lem:trigAexpB} into \eqref{lem:hharmonic} we deduce
 \begin{equation}\label{lem:trigexpconditions}
  \begin{aligned}
 & \l{0}=-\s{0} \text{ \& } \l{1}^2 + \l{2}^2 = 4\s{1}\s{2}\\
 & \text{ along with } \\
 & \s{1}>0 ,\ \s{2}>0
  \end{aligned}
 \end{equation}

  \item Conversely, substituting \eqref{lem:expAtrigB} into \eqref{lem:hharmonic} we deduce
 \begin{equation}\label{lem:exptrigconditions}
  \begin{aligned}
 & \l{0}=-\s{0} \text{ \& } \s{1}^2 + \s{2}^2 = 4\l{1}\l{2}\\
 & \text{ along with } \\
 & \l{1}>0 ,\ \l{2}>0
  \end{aligned}
 \end{equation}
\end{enumerate}
\end{subequations}
We now require a harmonic function $\gamma$ such that
\begin{equation}\label{lem:gammasquaredAB}
    \gamma_{a}^2+\gamma_{b}^2=e^{-2\tilde{h}}=A(a)+B(b),
\end{equation}
where $\tilde{h}$ is obtained from one of \eqref{lem:polynomialAB}--\eqref{lem:expAtrigB} subject to the corresponding conditions \eqref{lem:polynomialconditions}--\eqref{lem:exptrigconditions}. We note that
once one harmonic function $\gamma$ satisfying equation \eqref{lem:gammasquaredAB} is obtained, then all other solutions are known (see \cite{CS2010} for further discussion).

We now require a harmonic function $\gamma(a,b)$ satisfying \eqref{lem:gammasquaredAB},
where $A(a)$ and $B(b)$ belong to one of the three categories \eqref{lem:polynomialAB}--\eqref{lem:expAtrigB}.

\subsection{Case 1: Polynomial forms}

 We require a harmonic function $F(z)=\gamma(a,b)+i\delta(a,b)$ satisfying \eqref{lem:gammasquaredAB}, which by the Cauchy-Riemann equations may be written according to $\abs{F^{\prime}(z)}^{2}=A(a)+B(b)$.   As such we propose a harmonic function
\begin{equation}\label{eq:gammaharmonicpolynomial}
 F(z)= \sum_{n=0}^{N}\omega_{n}z^{n},
\end{equation}
where $\omega_{n}$ for $n\in\{0,1,2,\ldots,N\}$ are complex valued constants. Differentiating with respect to $z$, equation \eqref{lem:gammasquaredAB} may be written according to
\begin{equation}
 \abs{F^{\prime}(z)} = \left|\sum_{n=1}^{N}n\omega_{n}z^{n-1}\right|^{2}.
\end{equation}
 However, since  \eqref{lem:gammasquaredAB} must be satisfied, it follows that $2(N-1)\leq2$, in which case $N\leq 2$, which in turn implies
 \begin{multline}\label{eq:gammacoefficientspolynomial}
  \left|\omega_{1}\right|^2 + 4\re{\omega_{1}\bar{\omega}_{2}}a + 4\im{\omega_{1}\bar{\omega}_{2}}b + 4\left|\omega_{2}\right|^2(a^2+b^2)
  \\ = \l{0}+\s{0} + \l{1}a+\s{1}b + \l{2}a^2+\s{2}b^2.
 \end{multline}
It is immediately clear that $\l{2}=\s{2}$ as required, with $\l{2}=4\left|\w{2}\right|^2$, while we also observe
\begin{equation}
\omega_{1}=\frac{\left(\l{1}+i\s{1}\right)\w{2}}{\l{2}},
\end{equation}
which also ensures
\[4\l{2}\left(\l{0}+\s{0}\right) = \l{1}^2+\s{1}^2,\]
in line with  \eqref{lem:polynomialconditions}.

Without loss of generality we set $\omega_{0}=0$ which may be implemented by a translation of the $(x,y)$-coordinate system, while we may also assume $\omega_{2}=\frac{\sqrt{\lambda_{2}}}{2}$ which is always possible when multiplying $F(z)$ by an appropriate constant phase factor (cf. \cite{CS2010}). Thus the analytic function $F(z)$ may be explicitly written as
\begin{align}\label{eq:polynomialF}
F(z) &= \left[\frac{1}{2\sqrt{\lambda_{2}}}\left(\l{1}a-\s{1}b\right) + \frac{\sqrt{\l{2}}}{2}\left(a^2-b^2\right)\right] + i\left[\frac{1}{2\sqrt{\l{2}}}\left(\s{1}a+\l{1}b\right) + \sqrt{\l{2}}ab\right] \nonumber\\
     &=\gamma(a,b) + i \delta(a,b).
\end{align}

\subsection{Case 2: Exponential forms}
We require a harmonic function $F(z)=\gamma(a,b)\\+i\delta(a,b)$ such that $\abs{F^{\prime}(z)^{2}}=A(a)+B(b)$, where $A(a)$ and $B(b)$ are of the form \eqref{lem:trigAexpB}. As such we propose
\begin{equation}
 F(z)=\w{0}+\w{1}e^{-i\kappa z}+\w{2}e^{i\kappa z},
\end{equation}
where $\kappa\in\mathbb{R}$ is a parameter whose value will be determined in the following. Imposing $\abs{F^{\prime}(z)}^2=A(a)+B(b)$, we deduce $\kappa=\frac{k}{2}$ and
\begin{equation}
\abs{\w{1}}^{2}=\frac{4\s{1}}{k} \qquad \abs{\w{2}}^2=\frac{4\s{2}}{k^2}
\end{equation}
while setting $\w{1}=\abs{\w{1}}e^{i\phi_{1}}$, $\w{2}=\abs{\w{2}}e^{i\phi_{2}}$ and $-\l{1}+i\l{2}=\rho e^{i\psi}$, we find
\begin{equation}
e^{i(\phi_{1}-\phi_{2})} = \frac{1}{2\sqrt{\s{1}\s{2}}}\left(-\l{1}+i\l{2}\right)
\end{equation}
from which it immediately follows
\begin{equation}
 \phi_{1}-\phi_{2}=\psi \qquad 4\beta_{1}\beta_{2}=\l{1}^2+\l{2}^2.
\end{equation}
We define
\begin{equation}
 a_{1}=\frac{2}{k}\phi_{1}\quad a_{2}=\frac{2}{k}\phi_{2},
\end{equation}
and so we may write the analytic function $F(z)$ according to
\begin{equation}
\begin{aligned}
 F(z) &= \frac{2\sqrt{\s{1}}}{k}e^{-i\frac{k}{2}\left(z-a_{1}\right)}+\frac{2\sqrt{\s{2}}}{k}e^{i\frac{k}{2}\left(z-a_{2}\right)} \\
 &= \frac{2\sqrt{\s{1}}}{k}e^{\frac{kb}{2}}\cos\left[\frac{k}{2}\left(a-a_{1}\right)\right] + \frac{2\sqrt{\s{2}}}{k}e^{-\frac{kb}{2}}\cos\left[\frac{k}{2}\left(a-a_{2}\right)\right]
\\&\quad - i \frac{2\sqrt{\s{1}}}{k}e^{\frac{kb}{2}}\sin\left[\frac{k}{2}\left(a-a_{1}\right)\right] +i \frac{2\sqrt{\s{2}}}{k}e^{-\frac{kb}{2}}\sin\left[\frac{k}{2}\left(a-a_{2}\right)\right]\\
&=\gamma(a,b)+i\delta(a,b).
\end{aligned}
\end{equation}
The remaining solution for $F(z)$ is obtained from this by swapping the roles of $a$ and $b$.

\section{Physical flows and harmonic maps}
Physical flows of the form \eqref{eq:diffeomorphism} are defined in terms of analytic functions according to $F(z)\!+\!G(\bar{z})\!=\!f(a,b)\!+\!ih(a,b)$, where $F$ and $G$ take one of three forms as given by \eqref{lem:formsofF} and where $\mu(b)\!=\!\abs{F^{\prime}(z)}^{2}\!-\!\abs{G^{\prime}(\bar{z})}^{2}$ must be independent of $a$. Furthermore, from equation \eqref{eq:Fmunuxi} we may then deduce an expression for $\nu(b)$ which is essential when calculating vorticity in the flow.

\subsection{Parabolic particle paths}
In this case we consider analytic functions $F(z)$ and $G(\z)$, both of the form \eqref{eq:polynomialF}, subject to the condition that the difference of the square moduli of their derivatives is independent of $a$. We let
\begin{subequations}
\begin{align}
\label{eq:polynomialF1} F(z) &= \omega_{0} + \omega_{1}z + \omega_{2}z^2 \\
\label{eq:polynomialF2} G(\z) &= \zeta_{0} + \zeta_{1}\z + \zeta_{2}\z^{2}
\end{align}
\end{subequations}
in which case
\begin{multline}\label{eq:polynomialdifference}
\abs{F^{\prime}(z)}^2- \abs{G^{\prime}(\z)}^2 = \abs{\omega_{1}}^2-\abs{\zeta_{1}}^2 + 4\re{\omega_{1}\bar{\omega}_{2}-\zeta_{1}\bar{\zeta}_{2}}a\\
+ 4\im{\omega_{1}\bar{\omega}_{2}+\zeta_{1}\bar{\zeta}_{2}}b
+4\left(\abs{\omega_{2}}^2-\abs{\zeta_{2}}^2\right)\left(a^2+b^2\right).
\end{multline}
Given that this difference is required to be independent of $a$, it follows immediately that $\abs{\omega_{2}}=\abs{\zeta_{2}}$, which in turn also ensures
\begin{subequations}
\begin{align}
l_2^2+m_2^2&=L_2^2+M_2^2 \\
l_1l_2+m_1m_2&=L_1L_2+M_1M_2
\end{align}
\end{subequations}
where we introduce the notation
\begin{equation}\label{eq:cartesianparameters}
\begin{rcases}
\omega_{n} = l_n+im_n\\
\zeta_{n} = L_n+iM_n
\end{rcases}\text{ for }n\in\{0,1,2\}.
\end{equation}
Thus it follows that $\mu(b)=\abs{F^{\prime}(z)}^2- \abs{G^{\prime}(\z)}^2$ is given by
\begin{equation}\label{eq:polynomialmu}
\mu(b) = \abs{\omega_{1}}^2 - \abs{\zeta_{1}}^2 + 4\im{\omega_{1}\bar{\omega}_{2}  + \zeta_{1}\bar{\zeta}_{2}}b.
\end{equation}
It follows from equations \eqref{eq:Fmunuxi}, \eqref{eq:polynomialF1} and \eqref{eq:cartesianparameters}  that
\begin{equation}
  \xi(a) = 4\left(l_{1}l_{2}+m_{1}m_{2}\right)a + 4\left(l_{2}^2+m_{2}^2\right)a^2 \\
\end{equation}
Using equations \eqref{eq:complexmap} and \eqref{eq:polynomialF1}--\eqref{eq:polynomialF2} the particle trajectories within the flow are found to be given by
\begin{align}
 \label{eq:polynomialf}f(a,b) &=\! \left(L_{1}\!+\!l_{1}\right)a \!+\! \left(M_{1}\!-\!m_{1}\right)b \!+\! \left(L_{2}\!+\!l_{2}\right)\left(a^2-b^2\right) \!+\! 2\left(M_{2}\!-\!m_{2}\right)ab\\
 \label{eq:polynomialh}h(a,b) &=\! \left(m_{1}\!+\!M_{1}\right)a \!+\! \left(L_{1}\!-\!l_{1}\right)b \!+\! \left(m_{2}\!+\!M_{2}\right)\left(a^2\!-\!b^2\right) \!+\! 2\left(l_{2}\!-\!L_{2}\right)ab
\end{align}
where for convenience we have set $\zeta_{0}+\omega_0=0$, which is physically realised by  a translation of the $(x,y)$--plane.
\begin{figure}[h!]
 \centering
 \includegraphics[width=\textwidth]{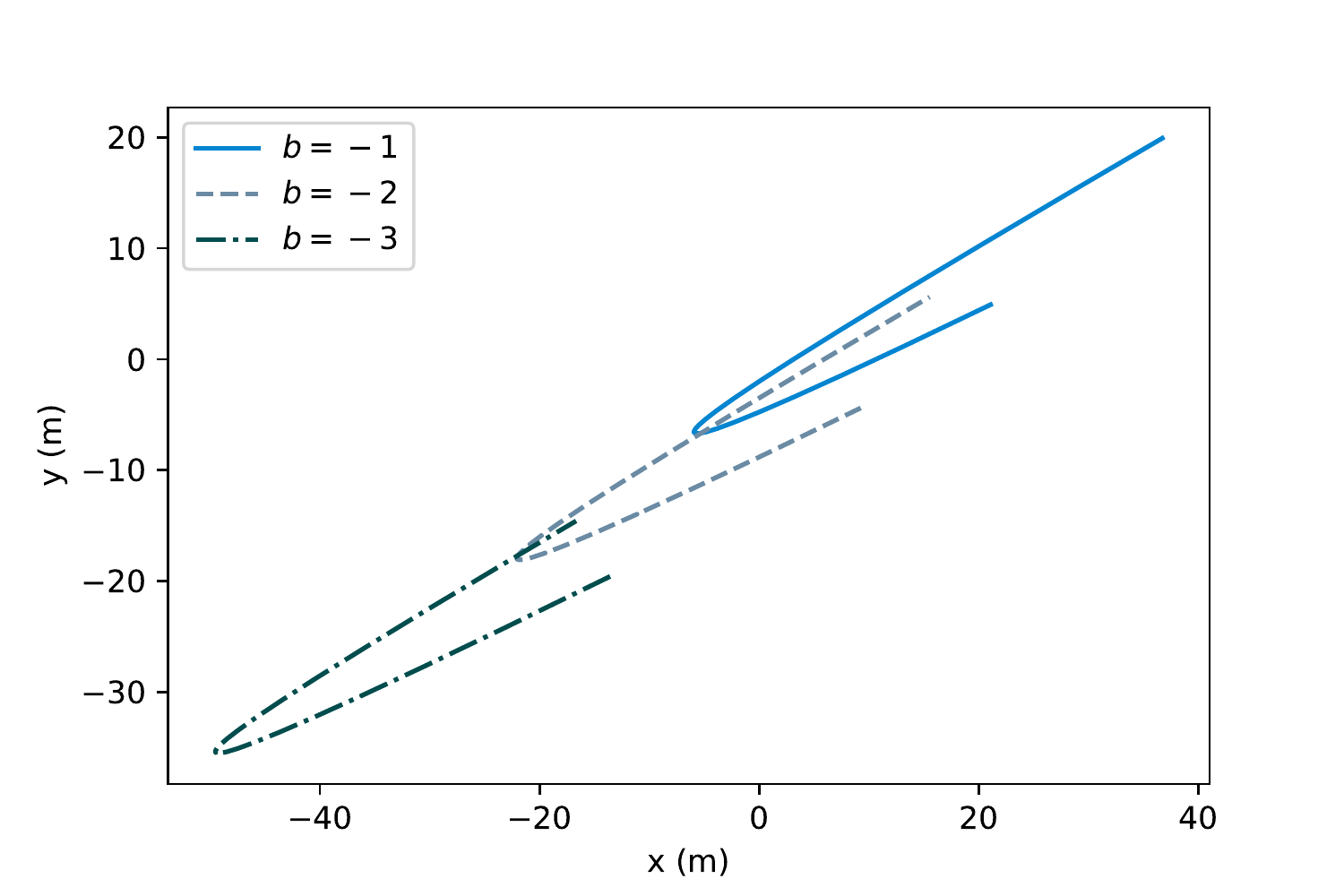}
 \caption{The particle trajectories given by \eqref{eq:polynomialf}--\eqref{eq:polynomialh} with $\omega_{1}=1+2i$, $\omega_{2}=3+i$, $\zeta_{1}=4+2i$ and $\zeta_{2}=2.5+1.94i$.}
\end{figure}

\subsubsection{Vorticity,  geophysical corrections and pressure distribution}
In general, the vorticity in two-dimensional flows is given by $\omega(x,y)=v_{x}-u_{y}$, and so from equations \eqref{eq:jacobianmatrix}--\eqref{eq:jacobian} we may write
\begin{equation}
 \begin{bmatrix}
  \partial_{x} \\ \partial_{y}
 \end{bmatrix}
= \frac{1}{\mu(b)}
\begin{bmatrix}
 h_{b} & -h_{a} \\ -f_{b} & f_{a}
\end{bmatrix}
\begin{bmatrix}
 \partial_{a} \\ \partial_{b}
\end{bmatrix}
\end{equation}
in which case the vorticity is given by
\begin{equation}\label{eq:vorticity}
  \omega(b) = v_{x} - u_{y}
                          = \frac{c}{\mu(b)}\left(f_{b}f_{aa}+h_{b}h_{aa}-f_{a}f_{ab}-h_{a}h_{ab}\right)
                         =\f-\frac{\nu^{\prime}(b)}{c\mu(b)}.
\end{equation}
as follows from \eqref{eq:nuprime}.  In contrast to the analogous expression for vorticity as found in \cite{CS2010}, we see that vorticity in the flow also has a geophysical contribution $2\Omega$ in the equatorial region. Nevertheless it remains true in this geophysical context that vorticity is constant along streamlines of the flow.  Combined with equation \eqref{eq:Fmunuxi} we then see that
\begin{equation}\label{eq:omega}
\omega(b)=4\Omega-c\frac{\partial_{b}\left(\abs{F^{\prime}(z)}^2+\abs{G^{\prime}(\z)}^2\right)}{\abs{F^{\prime}(z)}^2-\abs{G^{\prime}(\z)}^2}
\end{equation}
 In the case of hydrodynamic flows derived from polynomial pairs $F(z)$ and $G(\z)$, we find the vorticity in the flow is given by
 \begin{equation}
 \omega(b) = 4\Omega - c\frac{16\abs{\omega_{2}}^2b + 4\im{\omega_{1}\bar{\omega}_{2}-\zeta_{1}\bar{\zeta}_{2}}}{ \abs{\omega_{1}}^2- \abs{\zeta_{1}}^2 +4\im{\omega_{1}\bar{\omega}_{2}+\zeta_{1}\bar{\zeta}_{2}}b}.
\end{equation}
 As is clear, the vorticity is constant along curves of constant $b$, which is to say, vorticity is constant along streamlines of the flow (see \cite{Con2011} for further discussion of vorticity conservation along streamlines in 2-dimensional gravity driven flows in the absence of geophysical corrections).

In line with the boundary conditions in \eqref{eq:surfacecondition} we note that the free surface of a flow is a streamline where the pressure distribution is constant, thus allowing one to determine which pair $F$ and $G$ correspond to free boundary flows. In the current context  we consider a simple example where $\omega_{1}=l_1$, $\omega_{2}=im_{2}$, $\zeta_{1}=L_{1}$, $\zeta_{2}=im_{2}$ and whose Jacobian is
\[\mu(b)=\left(l_{1}-L_{1}\right)\left(l_{1}+L_{1}-4m_{2}b\right)\]
which is clearly non-negative for all $b\geq b_{crit}=\frac{l_{1}+L_{1}}{4m_{2}}$ when $l_{1}>L_{1}$ or $b<b_{crit}$ when $l_{1}<L_{1}$. In addition the hydrodynamic pressure distribution is given by
\begin{multline}
 P(a,b)=\frac{c^2}{2}\left(L_{1}-l_{1}\right)^{2} +  \left(L_{1}-l_{1}\right)\left(g+4m_{2}c^2+2\Omega c(L_{1}+l_{1})\right)b\\
               + 2m_{2}\left(g+4m_{2}c^2 + 2\Omega c\right)b^2 - 2m_{2}\left(g+4m_{2}c^2\right)a^2 .
\end{multline}
In contrast to \cite{CS2010} (where $\Omega=0$) we see that we may have a non-constant $P(a,b)$ which is independent of $a$ along fixed streamlines when $g+4m_{2}c^2=0$. This is only possible when $m_{2}<0$ and $c=\pm\sqrt{-\frac{g}{4m_{2}}}$, in which case the free surface $b=b_{0}$ is given by the solution of the quadratic equation
\begin{equation}\label{eq:parabolicpressure}
 \frac{c^2}{2}(L_{1}-l_{1})^2 + 2\Omega c\left(L_{1}^{2} - l_{1}^{2}\right)b + 4\Omega cm_{2}b^2 = P_{atm}.
\end{equation}
to give
\begin{equation}
 b_0=(l_{1}-L_{1})\left[b_{crit}\pm\sqrt{b_{crit}^2 - \frac{1}{4m_{2}\Omega c}\left(\frac{c^2}{2}-\frac{P_{atm}}{\left(L_{1}-l_{1}\right)^{2}}\right)}\right].
\end{equation}
Taking a typical wave speed  $c\!=\! \SI{1.4}{\meter \per \second}$ \cite{CRB2011} we see that we require $m_{2}\!\simeq\!\SI{-1.25}{\per\meter}$, while a typical atmospheric pressure at sea-level is $P_{atm}\!=\! \SI{101.325}{\square\meter\per\square\second}$ (after re-scaling by $\rho$) .
Moreover, we must choose values  $l_{1}\!>\!L_{1}$ such that the roots above are real and at least one root satisfies $b_{0}\!>\!b_{crit}$. The values $l_{1}\!=\!12$  and $L_{1}\!=\!1$ are such a pair, which then yields a parabolic curve in the $(x,y)$-plane whose height decays by about $\SI{160}{\meter}$ at about $\SI{100}{\meter}$ away from the crest-line. As such, quadratic forms for $F$ and $G$ are not a realistic choice for describing free boundary flows, and so we interpret such parabolic particle trajectories as being valid in the interior of geophysical flows.

\subsection{Elliptical paths}
\begin{figure}[ht]
\centering
\includegraphics[width=\textwidth]{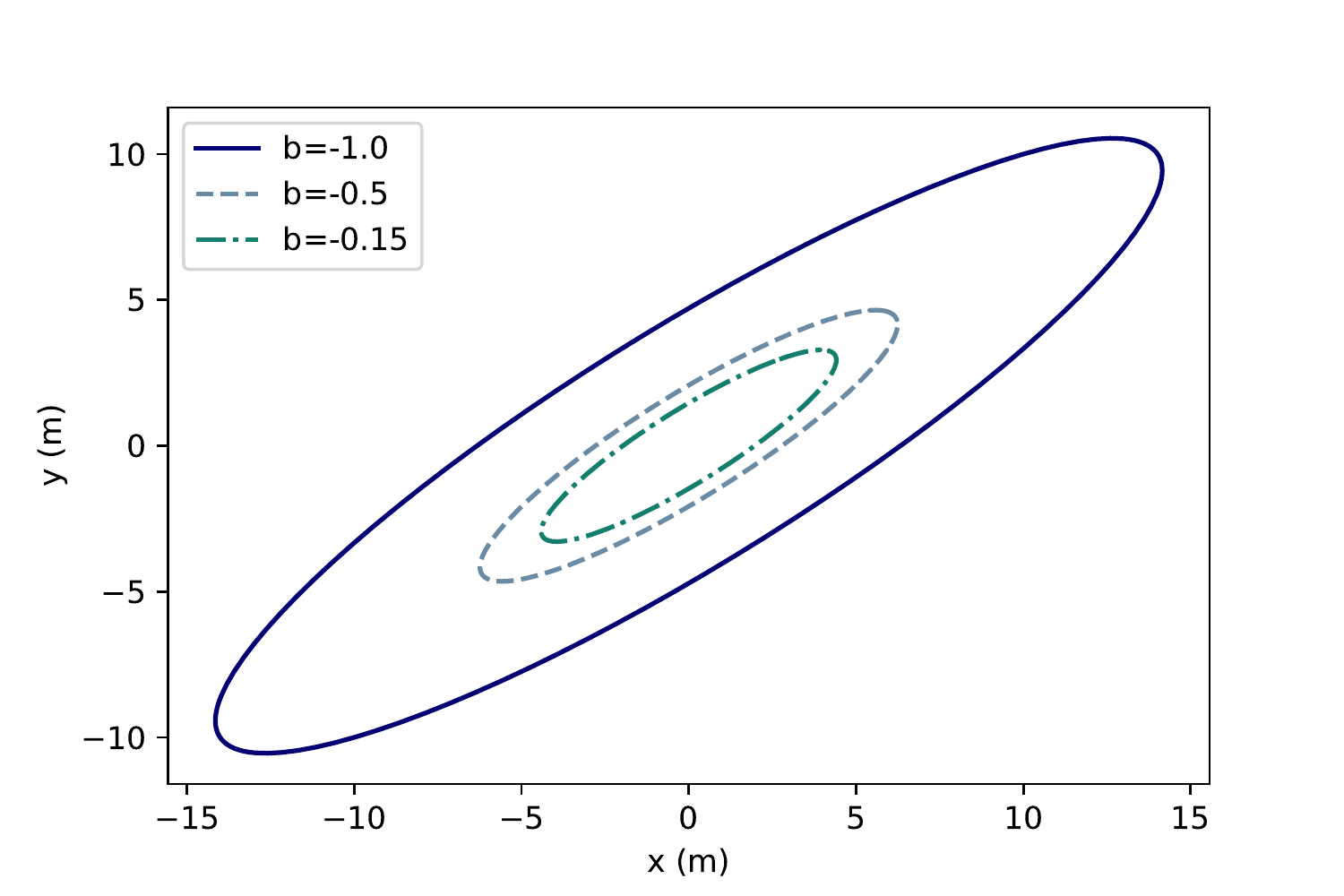}
\caption{Elliptical particle trajectories obtained from exponential profiles with complex parameters $\omega_{1}=1-2i$, $\omega_{2}=2+i$, $\zeta_{1}=1+i$ while the remaining parameter is $\kappa=\frac{\pi}{4}$ for $b\in\{-2.0,-1.0,-0.15\}$ with $b_{0}=-0.15$.}\label{fig:ellipse2}
\end{figure}

As a second example we consider particle paths obtained from a pair $F(z)$, $G(\z)$ both obtained from \eqref{lem:trigAexpB}, in which case
\begin{subequations}
\begin{align}
\label{eq:ellipticalF1} F(z) &=  \omega_{1}e^{-i\kappa z} + \omega_{2}e^{i\kappa z} \\
\label{eq:ellipticalF2} G(\z) &= \zeta_{1}e^{-i\kappa \z} + \zeta_{2}e^{i\kappa \z}
\end{align}
\end{subequations}
where we set $\omega_{0}=\zeta_{0}=0$ without loss of generality, with $\frac{k}{2}=\kappa\in\mathbb{R}$ a constant parameter.
With $\omega_{n}=l_{n}+im_{n}$ and $\zeta_{n}=L_{n}+iM_{n}$ for $n\in\{1,2\}$, we find the particle trajectory given by $F(z) + G(\z)=f(a,b) + i h(a,b)$ is
\begin{subequations}
\begin{align}
\label{eq:ellipticalf}f(a,b) &= \left[\left(l_{1} + L_{2}\right)e^{\kappa b} + \left(l_{2} + L_{1}\right)e^{-\kappa b}\right]\cos(\kappa a) \\
\nonumber &\quad +\left[\left(m_{1} - M_{2}\right)e^{\kappa b} - \left(m_{2} - M_{1}\right)e^{-\kappa b}\right]\sin(\kappa a), \\
\label{eq:ellipticalh}h(a,b) &= \left[ \left(m_{1} + M_{2}\right)e^{\kappa b} + \left(M_{1} + m_{2}\right)e^{-\kappa b}\right]\cos(\kappa a) \\
\nonumber &\quad +\left[\left(l_{2}-L_{1}\right)e^{-\kappa b} - \left(l_{1}-L_{2}\right)e^{\kappa b}\right]\sin(\kappa a),
\end{align}
\end{subequations}
in which case we see the fluid particles follow elliptical trajectories as illustrated in Figure \ref{fig:ellipse2}.

The Jacobian of the diffeomorphism \eqref{eq:diffeomorphism} resulting from the pair \eqref{eq:ellipticalF1}--\eqref{eq:ellipticalF2} is given by
\begin{equation}
\begin{aligned}
\mu(b) =  2\kappa^2\left[\re{\zeta_{1}\bar{\zeta}_{2}-\omega_{1}\bar{\omega}_{2}}\cos(ka) + \im{\zeta_{1}\bar{\zeta}_{2}-\omega_{1}\bar{\omega}_{2}}\sin(ka)\right] \\
 \kappa^2\left(\abs{\omega_{1}}^2-\abs{\zeta_{2}}^2\right)e^{kb} + \kappa^2\left(\abs{\omega_{2}}^2-\abs{\zeta_{1}}^2\right)e^{-kb},
\end{aligned}
\end{equation}
and so it follows at once that we require $\omega_{1}\bar{\omega}_{2} = \zeta_{1}\bar{\zeta}_{2}$ owing to the linear independence of $\cos(ka)$ and $\sin(ka)$. As such, we may rewrite the Jacobian for this flow according to
\begin{equation}
 \mu(b) = k^2\left(\abs{\omega_{1}\zeta_{1}} - \abs{\omega_{2}\zeta_{2}}\right)\sinh\left(k(b-b_0)\right),
\end{equation}
where we introduce the parameter $b_{0}=\frac{1}{k}\ln\abs{\frac{\zeta_{1}}{\omega_{1}}}$. Indeed, since the Jacobian is required to be non-negative we see that we must impose $b\in[b_{0},\infty)$, when $\abs{\omega_{1}\Omega_{1}}>\abs{\omega_{2}\zeta_{2}}$ or $b\in(-\infty,b_{0}]$, when $\abs{\omega_{1}\zeta_{1}}<\abs{\omega_{2}\zeta_{2}}$. In either case, we see that the diffeomorphism \eqref{eq:diffeomorphism} is valid only over a semi-infinite domain.

\begin{figure}[ht]
\centering
\includegraphics[width=\textwidth]{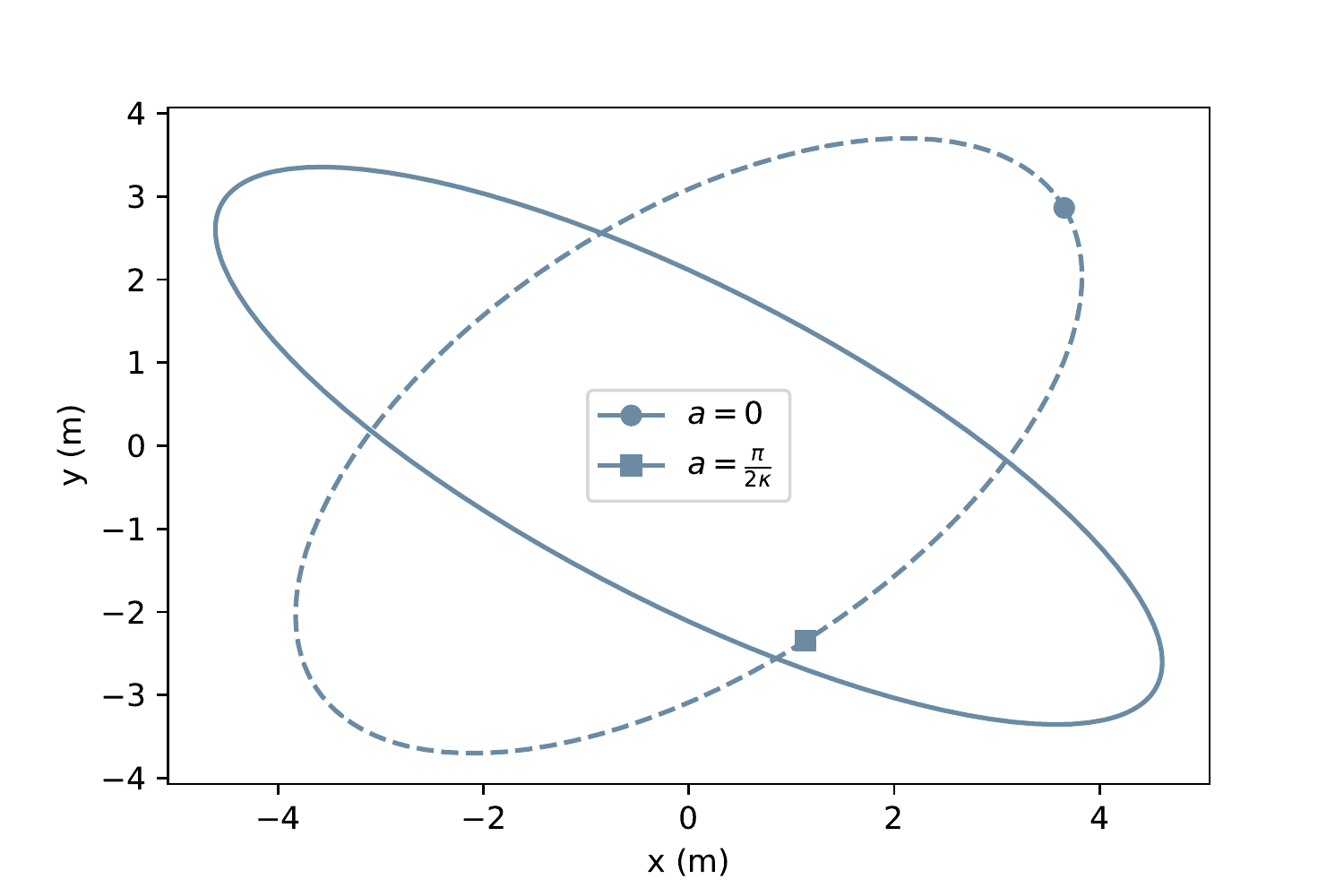}
\caption{An elliptical particle trajectory with $\omega_{1}=1+2i=\bar{\omega}_{2}$ and $\zeta_{1}=2-4i$ (solid curve) and a second trajectory with  $\omega_{1}=1-3i=\bar{\omega}_{2}$ and $\zeta_{1}=2+5i$ (dashed curve). The additional parameter is and $\kappa=\frac{\pi}{2}$.}\label{fig:ellipse1}
\end{figure}

The variable $b$  determines the length of the major and minor axes of the elliptical paths, all centred at the origin of the $(x,y)$-plane.  With reference to Figure \ref{fig:ellipse1} and equations \eqref{eq:ellipticalf}--\eqref{eq:ellipticalh} we see that the elliptical paths have semi-axes of length
\begin{equation}
\begin{aligned}
Q(b)=\sqrt{\left[\left(l_{1} + L_{2}\right)e^{\kappa b} + \left(l_{2} + L_{1}\right)e^{-\kappa b}\right]^2 + \left[ \left(m_{1} + M_{2}\right)e^{\kappa b} + \left(M_{1} + m_{2}\right)e^{-\kappa b}\right]^2}\\
R(b)=\sqrt{\left[\left(m_{1} - M_{2}\right)e^{\kappa b} - \left(m_{2} - M_{1}\right)e^{-\kappa b}\right]^2+\left[\left(l_{2}-L_{1}\right)e^{-\kappa b} - \left(l_{1}-L_{2}\right)e^{\kappa b}\right]^2},
\end{aligned}
\end{equation}
(for comparison see \cite{CS2010} where in the simpler setting $\omega_{n}$ and $\zeta_{n}$ were restricted to real values). This solution contrasts markedly with the well known elliptical paths one encounters in linearised irrotational Stokes waves over a flat bed, whose semi-major and semi-minor axes decrease with depth (see \cite{Con2011, MT2013} and references therein). As illustrated in Figure \ref{fig:ellipse2} where $\abs{\omega_{1}\zeta_{1}}<\abs{\omega_{2}\zeta_{2}}$, the diffeomorphism \eqref{eq:ellipticalf}--\eqref{eq:ellipticalh} is valid in the semi-infinite domain $(a,b)\in\mathbb{R}\times(-\infty,b_{0}]$ with $b_{0}\approx\SI{-0.15}{\meter}$ with $Q(b)$ and $R(b)$ monotonically decreasing functions of $b$ in this domain, while the centre of each elliptical path remains fixed at the origin.

To determine if and when such trajectories correspond to free boundary flows, we must examines the pressure distribution function associated with the analytic functions \eqref{eq:ellipticalF1}--\eqref{eq:ellipticalF2}. As a simplifying example we consider parameters of the form $\omega_{n}=\zeta_{n}=l_{n}\in\mathbb{R}$ for $n\in\{1,2\}$, and so we find
\begin{subequations}
\begin{align}
 f(a,b) &= \left(l_{2}+l_{1}\right)\cosh(\kappa b)\cos(\kappa a) \\
 h(a,b) &= \left(l_{2}-l_{1}\right)\cosh(\kappa b)\sin(\kappa a) \\
 \mu(b)&=\frac{k^2}{2}\left(l_{1}^2-l_{2}^2\right)\sinh(kb) \\
 \nu(b)&=\frac{c^2k^2}{2}\left(l_{1}^2 + l_{2}^2\right)\cosh(kb)+\Omega kc\left(l_{1}^2-l_{2}^2\right)\sinh(kb)\\
 \xi(a)&=\frac{c^2k^2}{2}l_{1}l_{2}\cos(ka),
\end{align}
\end{subequations}
from which it follows
\begin{multline}
 P(a,b) = \frac{k^2c^2}{2}\left(l_{1}^{2}+l_{2}^2\right)\cosh(kb) + \Omega kc\left(l_{1}^2-l_{2}^2\right)\sinh(kb)\\
 -\frac{k^2c^2}{8}\left[l_{1}^{2} + l_{2}^{2} + 2l_{1}l_{2}\cos(ka)\right]\cosh^2(\kappa b) \\
 + g\left(l_{1}-l_{2}\right)\cosh(\kappa b)\sin(\kappa a)
\end{multline}
It is clear that there is no fixed value of $b$ which makes $P(a,b)$ constant, and as such there is no streamline in the flow subject to constant pressure meaning there is no free boundary associated with this class of flows.

Given arbitrary parameters $\omega_{n}, \ \zeta_{n}$, we find the vorticity is given by
\begin{equation}
 \omega(b) = 4\Omega - 2ck\frac{\abs{\omega_{1}\zeta_{1}}-\abs{\omega_{2}\zeta_{2}}}{\abs{\omega_{1}\zeta_{1}}+\abs{\omega_{2}\zeta_{2}}},
\end{equation}
and so the vorticity is constant throughout these flows. Moreover, it is clear that with an appropriate choice of parameters $\omega_{n}$ and $\zeta_{n}$ the vorticity may actually vanish throughout the flow.

\subsection{Free boundary flows}
Thus far we have considered flows obtained when $F(z)$ and $G(\z)$ are both of similar form, i.e. either both polynomial form or both exponential form. However, under specific circumstances flows are allowed in which $F$ and $G$ may assume different forms. In this example we consider flows derived from the pair
\begin{subequations}
\begin{align}
 F(z)  &= \omega_{0} + \omega_{1}z + \omega_{2}z^2 \\
 G(\z) &= \zeta_{0} + \zeta_{1}e^{-i\kappa z} + \zeta_{2} e^{i\kappa z},
\end{align}
\end{subequations}
where $\kappa=\frac{k}{2}\in\mathbb{R}$ and without loss of generality we may set $\omega_{0}=\zeta_{0}=0$. As always, we require $\mu(b)=\abs{\dF}^2-\abs{\dG}^2$ which is explicitly given by
\begin{multline}
 \mu(b) = \abs{\omega_{1}}^2 + 4\re{\omega_{1}\bar{\omega}_{2}}a+4\im{\omega_{1}\bar{\omega}_{2}}b + 4\abs{\omega_{2}}^2\left(a^2+b^2\right) \\
 -\frac{k^2}{4}\left[\abs{\zeta_{1}}^2e^{-kb} + \abs{\zeta_{2}}^2e^{kb} - 2\re{\zeta_{1}\bar{\zeta}_{2}}\cos(ka) -  2\im{\zeta_{1}\bar{\zeta}_{2}}\sin(ka)\right]
\end{multline}
from which we deduce
\begin{equation}
 \omega_{2}=0 \text{ and } \zeta_{1}=0 \text{ or } \zeta_{2}=0.
\end{equation}
Physically, the particle velocity must remain bounded as $b\to-\infty$, in which case the condition $\zeta_{1}=0$ must be imposed if the solutions are to remain physically relevant.

The explicit solution is given by
\begin{subequations}
\begin{align}
 \label{eq:trochoidalf} f(a,b) &= l_{1}a-m_{1}b +L_{2}e^{\kappa b}\cos(\kappa a) - M_{2}e^{\kappa b}\sin(\kappa a) \\
\label{eq:trochoidalh} h(a,b) &= m_{1}a+l_{1}b +M_{2}e^{\kappa b}\cos(\kappa a) +L_{2}e^{\kappa b}\sin(\kappa a),
\end{align}
\end{subequations}
where we introduce $\omega_{1}=l_{1}+im_{1}$ and $\zeta_{2}=L_{2}+iM_{2}$. When we also impose the condition $m_{1}=L_{2}=0$, we obtain a class of trochoidal solutions first discovered by Gerstner \cite{Ger1809} and later rediscovered by Rankine \cite{Ran1863}. With $\zeta_{1}=\omega_{2}=0$, we find that the Jacobian of the flow is given by
\begin{equation}
 \mu(b) = \abs{\omega_{1}}^2-\kappa^2\abs{\zeta_{2}^2}e^{2\kappa b},
\end{equation}
and is non-negative in the semi-infinite interval $b\in(-\infty,b_{0}]$ and vanishes along the critical streamline
\begin{equation}
 b_{crit}=\frac{1}{\kappa}\ln\abs{\frac{\omega_{1}}{\kappa\zeta_{2}}}.
\end{equation}
In fact it is known that the map \eqref{eq:trochoidalf}--\eqref{eq:trochoidalh} with $m_{1}=L_{2}=0$ is a smooth diffeomorphism of the domain $(a,b)\in\mathbb{R}\times(-\infty,b_0)$ into the domain $(x,y)\in\mathbb{R}\times(-\infty,\eta_{0}(x))$, where the surface $y=\eta(x)$ is the curve $\left(f(a,b_{crit}), h(a,b_{crit})\right)$ for $a\in\mathbb{R}$ (see \cite{Con2001} for a proof).

\begin{figure}[ht]
\centering
\includegraphics[width=\textwidth]{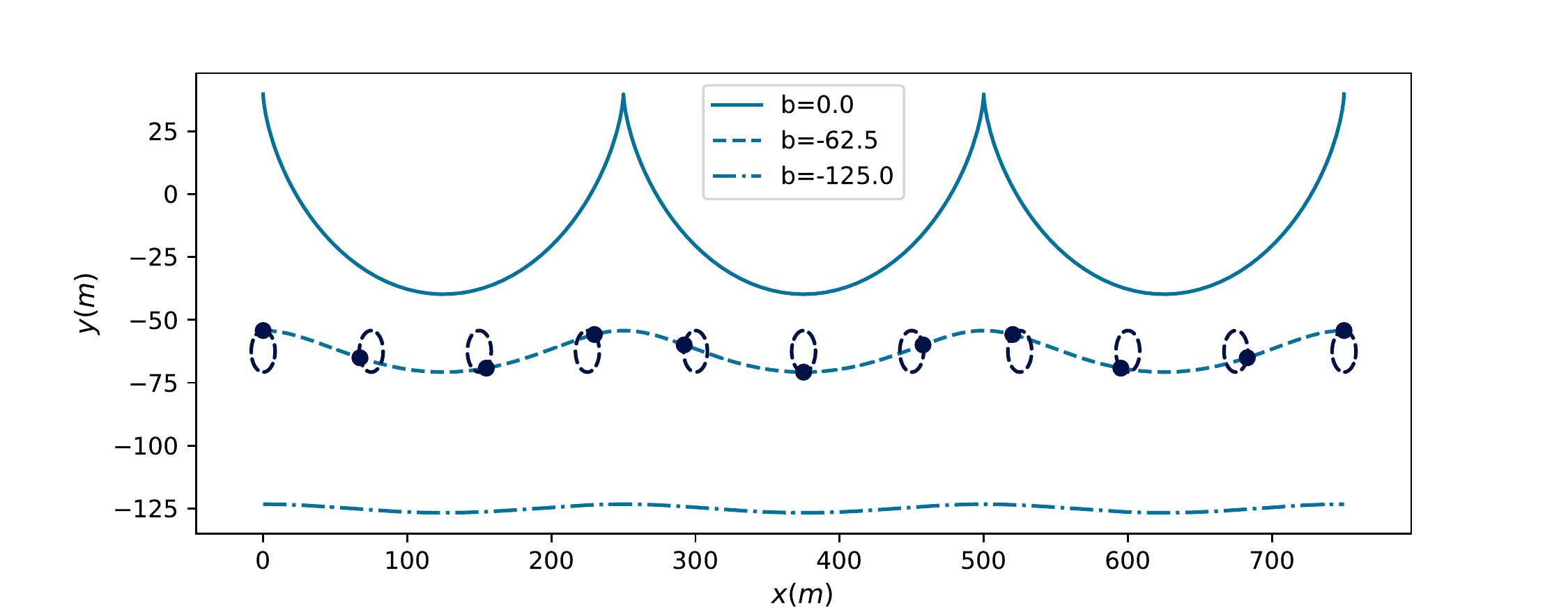}
\caption{A cycloid (solid curve) and trochoids (dashed curve) described by the map \eqref{eq:trochoidalf}--\eqref{eq:trochoidalh} wave with $\omega_{1}=1$ and $\zeta_{2}=5i$ and $\kappa=\frac{\pi}{10}$.}\label{fig:gerstner}
\end{figure}

Within the flow of the Gerstner-like solution (i.e. $m_{1}=L_{2}=0$) the particle trajectories are circular paths centered at $\left(l_{1}a, l_{1}b\right)$ with radii $M_{2}e^{\kappa b}$ (cf. Figure \ref{fig:gerstner}), with constant radii along a given streamline. The streamline $b=b_{crit}$ is a cycloid, that is to say it is the curve traced by a fixed point on a circle which rolls without slipping. On the other hand, for any streamline $b<b_{crit}$ the curve is a trochoid, which is the path traced by a fixed point on the interior of a rolling circle, with both curves illustrated in Figure \ref{fig:gerstner} below.

Since $\zeta_{1}=\omega_{2}=0$ in general, it is found that $\xi(a)=0$, irrespective of our choice for $\omega_{1}$ and $\zeta_{2}$. For an arbitrary choice of complex parameters $\omega_{1}$ and $\zeta_{2}$ we find
\begin{equation}
 \nu(b) = c^2\left[\abs{\omega_{1}}^2\left(1-\frac{2\Omega b}{c}\right) + \kappa\abs{\zeta_{2}}^2\left(\kappa - \frac{\Omega}{c}\right)e^{2\kappa b}\right].
\end{equation}
Thus we deduce the pressure distribution in a Gerstner type solution (i.e. $m_{1}=L_{2}=0$) is found to be
\begin{equation}
 P(a,b) =  \frac{c^{2} l_{1}^{2}}{2} - l_{1}\left(2 \Omega  c l_{1} + g\right)b  + \frac{M_{2}^2\kappa c}{2}\left(\kappa c -2 \Omega\right) e^{2 \kappa b} + M_{2}\left(\kappa c^{2} l_{1} - g \right)e^{\kappa b} \cos{\left(\kappa a \right)}.
\end{equation}
As always, free boundary solutions are permitted only when there exists at least one value of $b$ for which this expression for $P(a,b)$ is independent of $a$. Thus we see that Gerstner waves are indeed a class of free boundary solutions when the dispersion relation
\begin{equation}
 c=\pm\sqrt{\frac{g}{\kappa l_{1}}}.
\end{equation}
between the wave-velocity $c$ and the wave number $\kappa$ is satisfied. In this case the free boundary $b=b_0$ is obtained from
\begin{equation}
 P_{atm} =  \frac{c^2l_{1}^{2}}{2} - l_{1}\left(2c\Omega l_{1} + g\right)b + \frac{M_{2}^{2}\kappa c}{2}\left(\kappa c-2\Omega\right)e^{2\kappa b},
\end{equation}
which may be solved in terms of the Lambert $W$-function (cf. \cite{OLBC2010, VJC2000}). One definition of the Lambert $W$-function is given as the solution of the general relation $x=q+re^{sx} \Rightarrow x=q-\frac{1}{s}W\left(-rse^{qs}\right)$ where $q$, $r$ and $s$ may complex be constants,  (see \cite{KS2020, Lyo2019} for further applications of the Lambert function in the hydrodynamic setting).

\begin{figure}[ht]
\centering
\includegraphics[width=\textwidth]{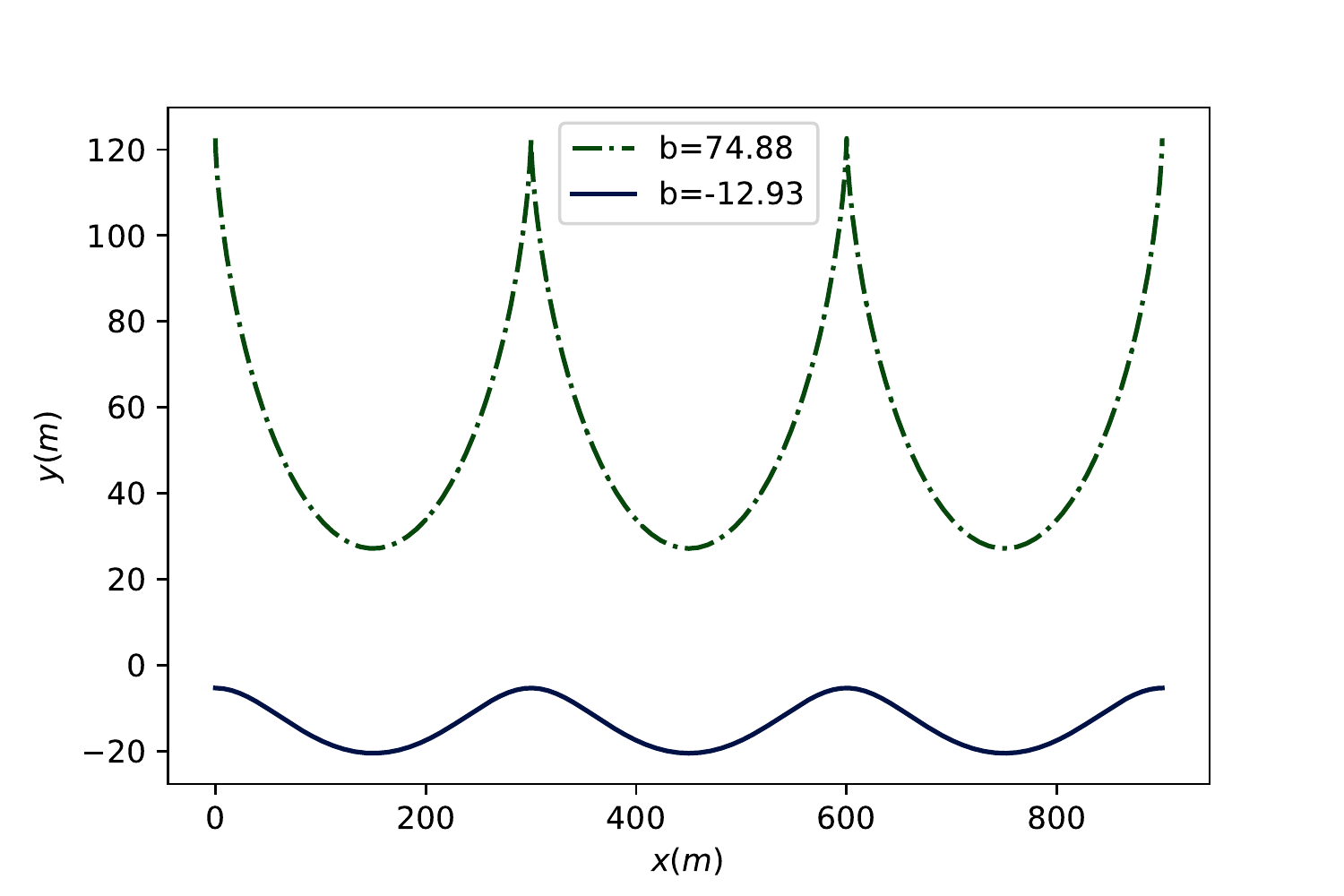}
\caption{The free surface $b=b_{0}=-\SI{12.93}{\meter}$ (solid curve) and the cycloid $b=b_{crit}=\SI{74.88}{\meter}$ when $l_{1}=1$ and $M_{2}=9.95$ with $P_{atm}=\SI{101.325}{\square\meter\per\square\second}$ and wavelength $\SI{300}{\meter}$.}\label{fig:gerstner2}
\end{figure}

The dispersion relation $c^2=\frac{g}{\kappa}$ is unaffected by geophysical effects in the vicinity of the equator (cf. \cite{Con2012}) and as such we choose $l_{1}=1$. The value $M_{2}=\frac{1}{\kappa}$ is often used  to describe Gerstner waves (see \cite{Con2011, Hen2008} for instance), however we note that the Lambert $W$--function is real valued only when its argument is greater than $-e^{-1}$ which modifies the value we may choose for $M_{2}$ in the current example. A typical wavelength in for surface gravity waves in the equatorial Pacific is of the order $\lambda=\SI{300}{\meter}$ \cite{Kins1965} giving an approximate wave-speed $c=\SI{22}{\meter\per\second}$, while the standard atmospheric pressure at sea-level is taken as $P_{atm}=\SI{101.325}{\square\meter\per\square\second}$ (when re-scaled by $\rho$). Using these parameters we find $M_{2}\lesssim9.969$. The critical surface and the free boundary of this flow are shown in Figure \ref{fig:gerstner2} when $M_{2}=9.95$.

In general, we find that the vorticity in the flow is given by
\begin{equation}
 \omega(b) = 4\Omega - c \frac{2\kappa^3\abs{\zeta_{2}}^2e^{2\kappa b}}{\abs{\omega_{1}}^2-\kappa^2\abs{\zeta_{2}}^2e^{2\kappa b}},
\end{equation}
for general complex parameters $\omega_{1}$ and $\zeta_{2}$.
This vorticity distribution is singular along the cycloid $b=b_{crit}$ and approaches $4\Omega$ as $b\to-\infty$. We see that for positive $c$ the vorticity may be negative on streamlines near the free surface $b=b_{0}$ and will change sign across the streamline
\begin{equation}
b^{\dag} = b_{crit} + \frac{1}{2\kappa}\ln\frac{2\Omega}{2\Omega+\kappa c}.
\end{equation}
In the limit $\Omega\to0$ we find $b^{\dag}\to-\infty$, in which case the vorticity is negative throughout the flow for positive $c$, and so we recover the analogous result for free boundary flows obtained in \cite{CS2010}.
\section*{Acknowledgements}
The author is grateful to the anonymous referees for several helpful suggestions.

\medskip
Received December 2021; revised January 2022; early access February 2022.
\medskip

\end{document}